\documentclass[amssymb,prd,aps,amsmath,twocolumn,superscriptaddress,nofootinbib,preprintnumbers,10pt]{revtex4-1}
\pdfoutput=1
\usepackage[utf8]{inputenc}
\usepackage{graphicx}
\usepackage{psfrag}
\usepackage{braket}
\usepackage[usenames,dvipsnames]{color}
\usepackage{epsfig,amsmath,amssymb,fancyhdr}   
\usepackage[normalem]{ulem}
\usepackage{amsmath}
\usepackage{slashed}
\usepackage{mathtools}
\usepackage[titletoc]{appendix}
\usepackage{datetime}
\usepackage{alltt} 
\usepackage{upgreek}
\usepackage{enumitem}
\usepackage{epstopdf} 
\usepackage{listings}
\usepackage{xargs}
\usepackage{capt-of}
\usepackage{ulem}
\usepackage{caption}
\usepackage{subcaption}

\usepackage{array}
\newcolumntype{L}[1]{>{\raggedright\let\newline\\\arraybackslash\hspace{0pt}}m{#1}}
\newcolumntype{C}[1]{>{\centering\let\newline\\\arraybackslash\hspace{0pt}}m{#1}}
\newcolumntype{R}[1]{>{\raggedleft\let\newline\\\arraybackslash\hspace{0pt}}m{#1}}

\def\ha{{\mathcal{H}}}

\def\kd{{\dot{\kappa}}}
\def\icg{{\dot{\mu}}}

\def\nn{{\nonumber}}
\def\ugc{{u_{\gamma\mathrm{DM}}}}

\def\gdm{{\gamma\mathrm{DM}}}
\def\lcdm{{$\Lambda$CDM}}

\begin{document}

\title{Is it Mixed Dark Matter or neutrino masses? }
\author{Julia Stadler}
\email{julia.j.stadler@durham.ac.uk}
\affiliation{Institute for Particle Physics Phenomenology, Durham University, South Road, Durham, DH1 3LE, United Kingdom}
\affiliation{Max Planck Institute for Extraterrestrial Physics, Giessenbachstrasse 1, 85748 Garching, Germany}

\author{C{\'e}line B{\oe}hm}
\affiliation{Institute for Particle Physics Phenomenology, Durham University, South Road, Durham, DH1 3LE, United Kingdom}
\affiliation{LAPTH, U. de Savoie, CNRS, BP 110, 74941 Annecy-Le-Vieux, France}
\affiliation{School of Physics, University of Sydney, Camperdown, NSW 2006, Australia}

\author{Olga Mena}
\affiliation{IFIC, Universidad de Valencia-CSIC, 46071, Valencia, Spain}

\preprint{}

\begin{abstract}
In this paper, we explore a scenario where the dark matter is a mixture of interacting and non interacting species. Assuming dark matter-photon interactions for the interacting species, we find that the suppression  of the matter power spectrum in this scenario can mimic that expected in the case of massive neutrinos. Our numerical studies include present limits from Planck Cosmic Microwave Background data, which render the strength of the   dark matter photon interaction unconstrained when the fraction of interacting dark matter is small. Despite the large entangling between mixed dark matter and neutrino masses, we show that future measurements from the Dark Energy Instrument (DESI) could help in establishing the dark matter and the neutrino properties simultaneously, provided that the interaction rate is very close to its current limits and the fraction of interacting dark matter is at least of $\mathcal{O}\left(10\%\right)$. However, for that region of parameter space where a small fraction of interacting DM coincides with a comparatively large interaction rate, our analysis highlights a considerable degeneracy between the mixed dark matter parameters and the neutrino mass scale.
\end{abstract}

\maketitle
\section{Introduction}
\label{sec: intro}
The standard Lambda Cold Dark Matter cosmological model (\lcdm) describes the observed angular power spectrum of the Cosmic Microwave Background (CMB) remarkably well. It is also very successful in predicting the Universe's large-scale-structure distribution (LSS). Yet, the particle physics nature of dark matter (DM) remains elusive.  A possible way to reveal the microscopic properties of DM and to unveil how collisionless DM needs to be to explain the observed Universe is to assume that DM is not a cold, collisionless fluid. 

Models of collisional DM have been  extensively studied in the literature. Namely, there have been models on DM-photon interactions in Refs.~\cite{Boehm:2000gq, Boehm:2001hm, Boehm:2004th, Sigurdson:2004zp, CyrRacine:2012fz, Dolgov:2013una,  Wilkinson:2013kia, Boehm:2014vja, Schewtschenko:2014fca, Schewtschenko:2015rno, Escudero:2015yka, McDermott:2010pa, Diacoumis:2017hff, Ali-Haimoud:2015pwa, Stadler:2018jin}, on DM-neutrinos in Refs.~\cite{Boehm:2000gq, Boehm:2004th, Mangano:2006mp, 2010PhRvD..81d3507S, Wilkinson:2014ksa, DiValentino:2017oaw, Ali-Haimoud:2015pwa}, on DM-baryon 
interactions in e.g. Refs.~\cite{Chen:2002yh, Dvorkin:2013cea, Dolgov:2013una, CyrRacine:2012fz, Prinz:1998ua} and on DM self-interactions in Refs.~\cite{Carlson:1992fn, deLaix:1995vi, Spergel:1999mh, Dave:2000ar, Boehm:2000gq, Boehm:2004th, Creasey:2016jaq, Rocha:2012jg, Kim:2016ujt, Huo:2017vef, Markevitch:2003at, Randall:2007ph}. Interactions of DM particles with a hypothetical dark radiation component have also been considered in e.g. Refs.~\cite{Das:2017nub, Kaplan:2009de, Diamanti:2012tg,  Buen-Abad:2015ova, Lesgourgues:2015wza, Ko:2017uyb, Escudero:2018thh, Das:2010ts, Archidiacono:2019wdp}. 

These DM interactions strongly impact the CMB fluctuations and the matter power spectrum below a certain cut-off scale, imposing strong limits on the cosmological epochs during which such a coupling can be efficient. If, however, the properties of only a fraction of DM differ from the common assumption of a cold, collisionless fluid, a more varied, subtle impact on the cosmological history arises. So have admixtures of cold and warm DM and their potential to resolve small scale discrepancies of \lcdm \cite{Bullock:2017xww} been studied in the recent literature \cite{Boyarsky:2008xj, Schneider:2018xba, Gariazzo:2017pzb, Diamanti:2017xfo, Anderhalden:2012qt}. Further works considered a fraction of DM interacting with neutrinos \cite{Serra:2009uu}, with dark radiation \cite{Chacko:2016kgg,Buen-Abad:2017gxg,Foot:2014uba} or a subdominant population of electrically charged massive particles \cite{Kamada:2017icv}. In several cases these scenarios are able to resolve or lessen tensions in current cosmological data sets, making them very attractive from a phenomenological perspective. In this paper we take a further step towards the general understanding of mixed DM scenarios and study the combination of a cold, collisionless DM component and a collisional DM component, which experiences elastic scattering off photons.

The damping of small-scale perturbations in this scenario has a rich and interesting phenomenology. In particular a small fraction of interacting DM causes a step-like suppression in the matter power spectrum below some characteristic scale -- very alike that produced by heavy neutrinos. Therefore, one may confuse those two scenarios when interpreting the data from galaxy surveys. Additional complications arise from the scale dependence of the halo bias introduced by neutrino masses, which will be important at the expected accuracy of future galaxy surveys \cite{Raccanelli:2017kht, Munoz:2018ajr, Vagnozzi:2018pwo, Giusarma:2018jei}. In this regard, we consider the Dark Energy Instrument (DESI)~\cite{Levi:2013gra, Aghamousa:2016zmz} and discuss to what extend and over which region of the parameter space its determination of the neutrino mass scale can be corrupted by the presence of mixed DM.

This paper is organised as follows. In Sec.~\ref{sec: model} we present the mixed-DM model. The implications of this scenario on the CMB and the current constraints from the Planck satellite, using the 2015 data likelihood release, are discussed in Sec.~\ref{sec:impact-cmb}. Section~\ref{sec: impact-pk} contains both a detailed description of the mixed-DM model's impact on LSS and a devoted forecast for the expected sensitivity of the future DESI galaxy survey on the model parameters. We draw our conclusions in Sec.~\ref{sec: conclusions}. 

\section{Model and implementation}
\label{sec: model}
We consider a scenario in which two additional heavy species, besides the Standard Model degrees of freedom, contribute to the universe's present matter density. One obeys the standard assumptions of Cold Dark Matter (CDM) and interacts only gravitationally, the second experiences interactions with photons ($\gdm$). Within the $\Lambda$CDM model, heavy neutrinos are often referred to as a second, hot DM component. However, in the context of the present work, we find it less ambiguous to reserve the term DM for the aforementioned CDM and $\gdm$ species, which are highly non-relativistic at recombination and thus behave as matter on all time and length scales of interest. For simplicity,  we shall assume that the elastic scattering cross section associated with the $\gdm$ component is independent of the DM energy and velocity and consequently it is described by a constant ($\sigma_\gdm$). In this model, the total amount of DM in the universe is the sum of both components, that is $\Omega_\mathrm{DM} = \Omega_\mathrm{CDM} + \Omega_\gdm$. The fraction of interacting DM is defined as $f_\gdm \equiv\Omega_\gdm/\Omega_\mathrm{DM}$.\\

The impact of the $\gdm$ component on the CMB is described by a single parameter: the ratio of the scattering cross section to the DM mass, parameterised as
\begin{equation}
\ugc = \frac{\sigma_\gdm}{\sigma_\mathrm{Th}}\left(\frac{m_\gdm}{100~\mathrm{GeV}}\right)^{-1}\,.
\end{equation}
In scenarios where all the DM is interacting (i.e. $f_\gdm = 1$) it was found, using the 2015 Planck data \cite{Adam:2015rua,Aghanim:2015xee,Ade:2015xua}, that $u_\gdm $ cannot exceed $u_\gdm \leq 2.25 \times 10^{-4}$ at $95\%$~CL \cite{Stadler:2018jin} due to the damping of the acoustic peaks at large multipoles, but there is no limit yet for scenarios with a smaller fraction of interacting DM.

The evolution of the CDM and $\gdm$ perturbations in the linear regime are described by two different sets of equations. Those concerning the CDM component are given in Ref.~\cite{Ma:1995ey}, while expressions for the $\gdm$ component where derived in Ref.~\cite{Stadler:2018jin}. 
The most important term for the purpose of the current analysis is the  velocity dispersion of the interacting component, $\theta_\gdm$. It has an additional scattering term with respect to CDM, and reads as
\begin{align}
\dot{\theta}_\gdm =& -\ha\theta_\gdm + c_\gdm^2k^2\delta_\gdm + k^2\psi \nn\\
&- S\icg\left(\theta_\gdm - \theta_\gamma \right)~,
\label{eq: theta-gdm}
 \end{align}
where $\phi$ and $\psi$ are the metric perturbations in Newtonian gauge, $\icg=a\,n_\gdm\,\sigma_\gdm$ is the $\gdm$ scattering rate, $c_\gdm$ the sound speed of the $\gdm$ component, the ratio $S=4\rho_\gamma/3\rho_\gdm$ ensures momentum conservation and $\ha= aH$. Our notation follows closely that of \cite{Ma:1995ey}. 

We note that the sound speed term in the RHS of Eq.~(\ref{eq: theta-gdm}) is present whenever DM particles are in thermal contact with the photon bath. Yet, this term changes the CMB ($P(k)$) predictions on observable scales only when  $m_\gdm$ (the mass of the interacting component) is smaller than $10~\mathrm{eV}$ ($1~\mathrm{GeV}$) ~\cite{Stadler:2018jin}.  Since the cosmology of very light DM particles can be very different from CDM (see for example Ref.~\cite{Marsh:2015xka}), we will  restrict our analysis to  the most CDM-like scenarios for now and assume $m_{\gdm}  \gtrsim 1~\mathrm{GeV}$  to neglect the subtle effect of the sound speed and focus on the main effect of mixed DM in what follows. 

The evolution of the photon perturbations is affected by both DM components through the gravitational potentials and by the DM-$\gamma$ interactions. The equation associated to the photon velocity dispersion thus reads
\begin{align}
\dot{\theta}_\gamma =& k^2\left(\frac{1}{4}\delta_\gamma-\sigma_\gamma\right) + k^2\psi + \kd\left(\theta_b-\theta_\gamma\right)\nn\\
 &+ \icg\left(\theta_\gdm-\theta_\gamma\right)\,.
\end{align}

Further, in the equations of the higher order multipoles $F_{\gamma,l}$ ($l\ge3$) and second Stokes parameter's multipoles $G_{\gamma,k}$ ($k=0,1,...$) each occurrence of $\kd$ in the source terms is replaced by $(\kd+\icg)$~\cite{Ma:1995ey,Stadler:2018jin}.

The evolution of neutrino and baryon perturbations is not modified by the introduction of an interacting DM component, except for the impact of the interacting component on the gravitational potentials. Their expressions can be found in \cite{Ma:1995ey}.

We now discuss the impact that these modifications have on the CMB spectra and on the matter power spectrum. All the modifications to the Boltzmann equations in the mixed-DM scenario have been implemented in the Boltzmann code CLASS\footnote{http://class-code.net/} (version 2.6) \cite{Blas:2011rf,Lesgourgues:2011re}.

\section{Impact on CMB spectra and parameter constraints}
\label{sec:impact-cmb}
\begin{figure}
	\includegraphics[]{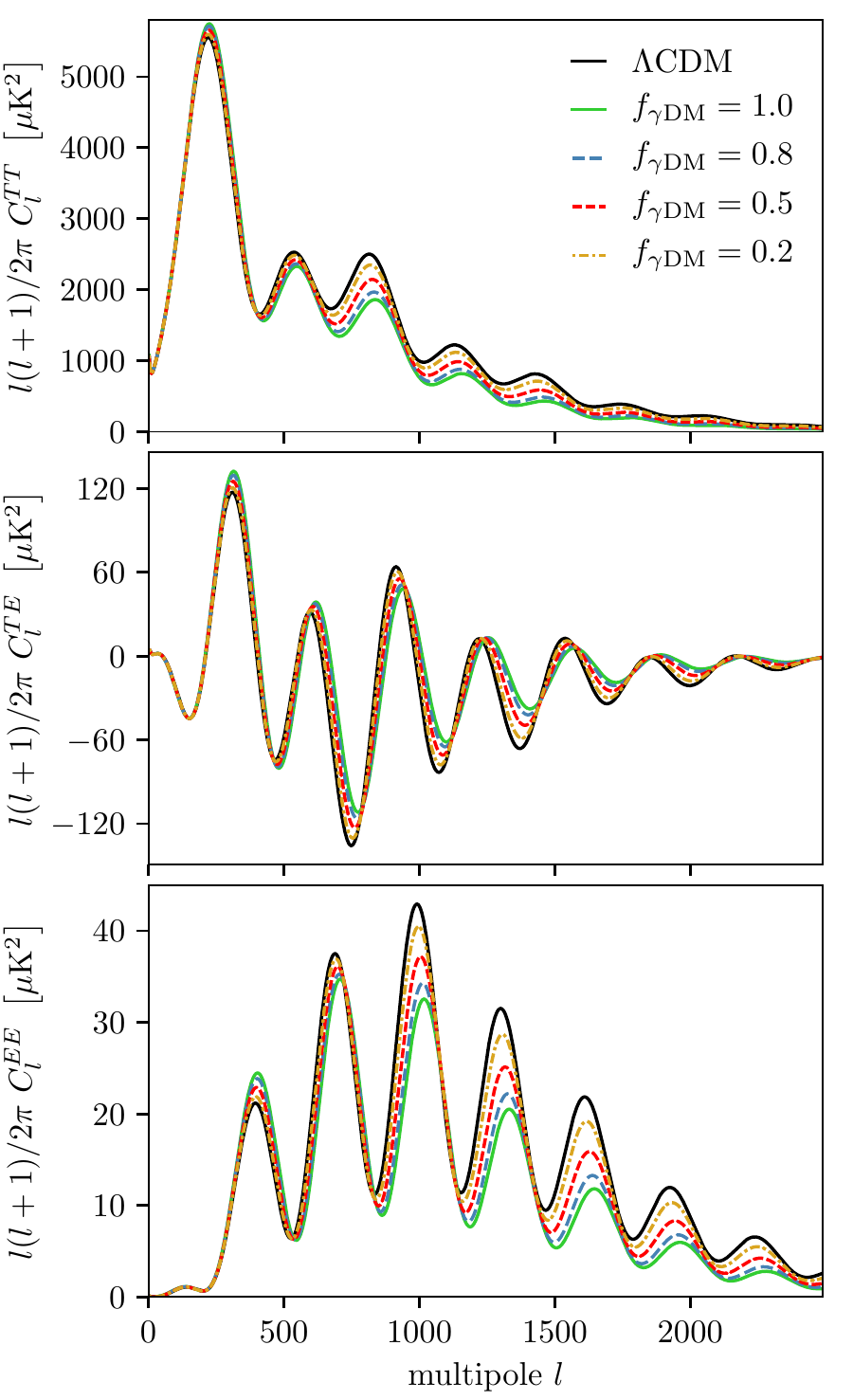}
	\caption{Impact of DM photon scattering on the CMB spectra for a cross section to mass ratio of $\ugc=0.01$ and several interacting DM fractions.}
	\label{fig: results-cls}
\end{figure}

An interacting DM component with $f_\gdm =1$ affects the CMB temperature and polarisation spectra via: \textit{(i)} an increase of the first acoustic peak caused by the decrease in the photon's diffusion length; \textit{(ii)} a reduction of all acoustic peaks due to collisional damping, and, \textit{(iii)} an overall shift of the Doppler peaks towards higher multipoles as a result of the decreased sound speed of the plasma~\cite{Wilkinson:2013kia,Boehm:2001hm}. Hence we expect that some of these features are also present in mixed DM scenarios. We compare in Fig.~\ref{fig: results-cls} the  temperature auto-correlation (TT), the E-mode  polarisation auto-correlation (EE) and the temperature E-mode cross correlation (TE) spectrum for \lcdm, pure-$\gdm$, and mixed-DM varying the fraction of interacting DM, $f_\gdm$. We have chosen a large cross section to mass ratio ($\ugc=0.01$) to enhance the effects. Mixed-DM has a similar effects on the CMB spectra as pure-$\gdm$, though less pronounced. 
As a result, the  TT, TE, and EE spectra obtained for mixed-DM are intermediate between the \lcdm~and the pure-$\gdm$ case. The $f_\gdm$ fraction essentially controls the interpolation between these two limits.

\begin{figure*}
	\includegraphics[width=1.0\textwidth]{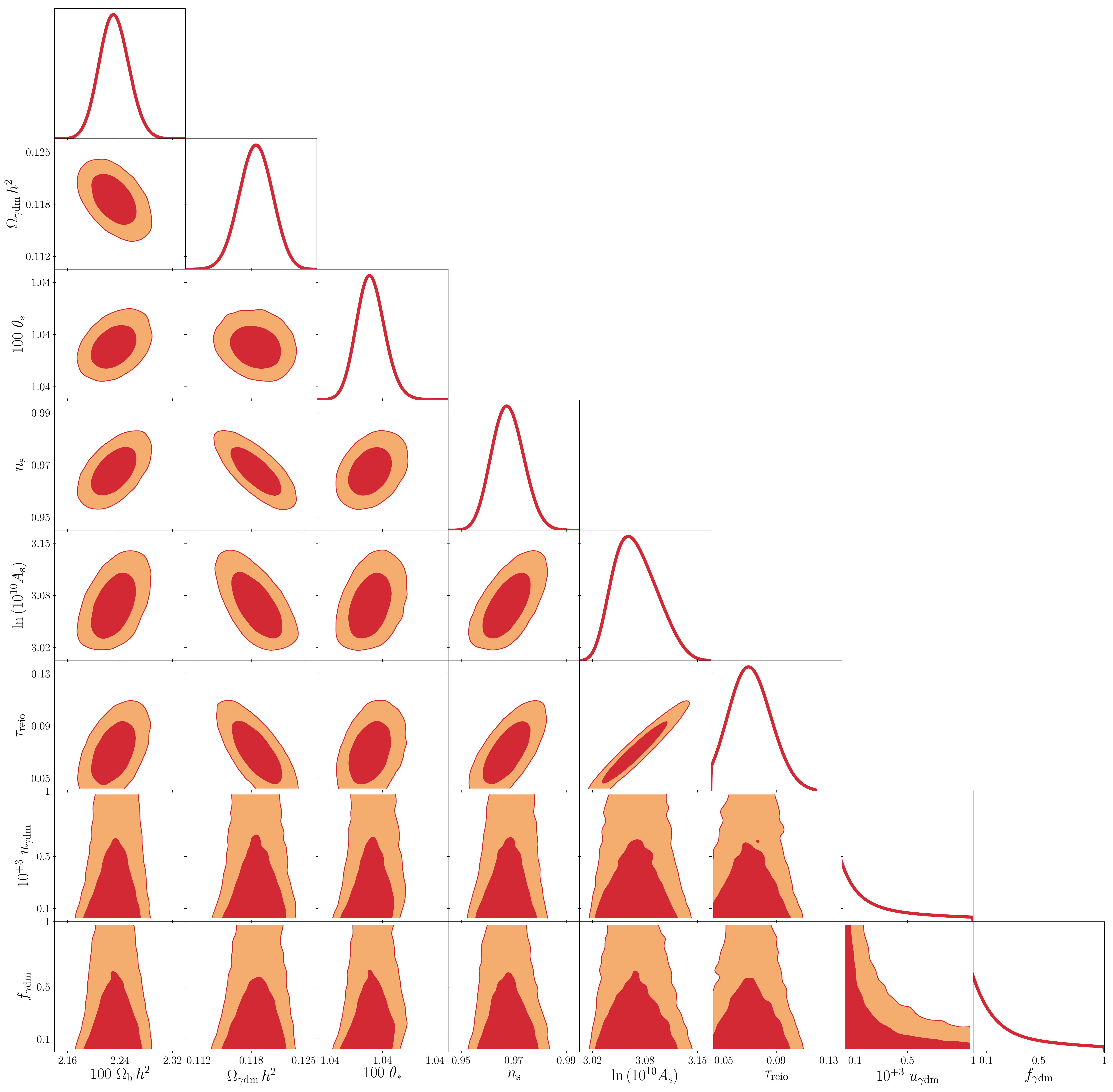}
	\caption{Two dimensional contours $68\%$ and $95\%$ CL contours and one dimensional posterior probabilities for the cosmological parameters of the mixed $\gamma$DM scenario.}
	\label{fig:mcmc}
\end{figure*}

In order to derive the constraints on mixed $\gamma$DM scenarios from the current CMB publicly available data, we shall exploit in the following measurements from the Planck 2015 data release~\cite{Adam:2015rua, Aghanim:2015xee}. We make use of the TT likelihood at high multipoles ($30\le \ell \le 2508$), the temperature and polarisation data at low multipoles  ($2\le \ell \le 29$) and the lensing likelihood. Additional parameters, such as those related to foreground contamination, calibration, and others have been marginalised over when deriving the final constraints, for which we use the Markov Chain Monte Carlo (MCMC) tool Monte Python~\cite{Audren:2012wb,Brinckmann:2018cvx}, interfaced with the Boltzmann solver CLASS~\cite{Blas:2011rf,Lesgourgues:2011re}. We assume flat priors on all cosmological and nuisance parameters. Figure~\ref{fig:mcmc} presents the results from our Monte Carlo analyses. Note the huge degeneracy between the interaction rate $u_\gdm$  and the interacting DM fraction $f_\gdm$ and how strong its anticorrelation is. In a further set of MCMC runs, we fix the fraction of interacting DM to a value smaller than one, which leaves $u_\gdm$ and the six $\Lambda$CDM parameters to be sampled. Considering $f_\gdm = 0.5\,,~0.1\,,~0.05$ we obtain as an upper limit for the interaction strength parameter at 95\% C.L. $4 \times 10^{-4}$, $3 \times 10^{-3}$ and $0.019$, respectively.
 
\section{Impact on the linear matter power spectrum: expectations from future galaxy surveys}
\label{sec: impact-pk}

\begin{figure}
\includegraphics[]{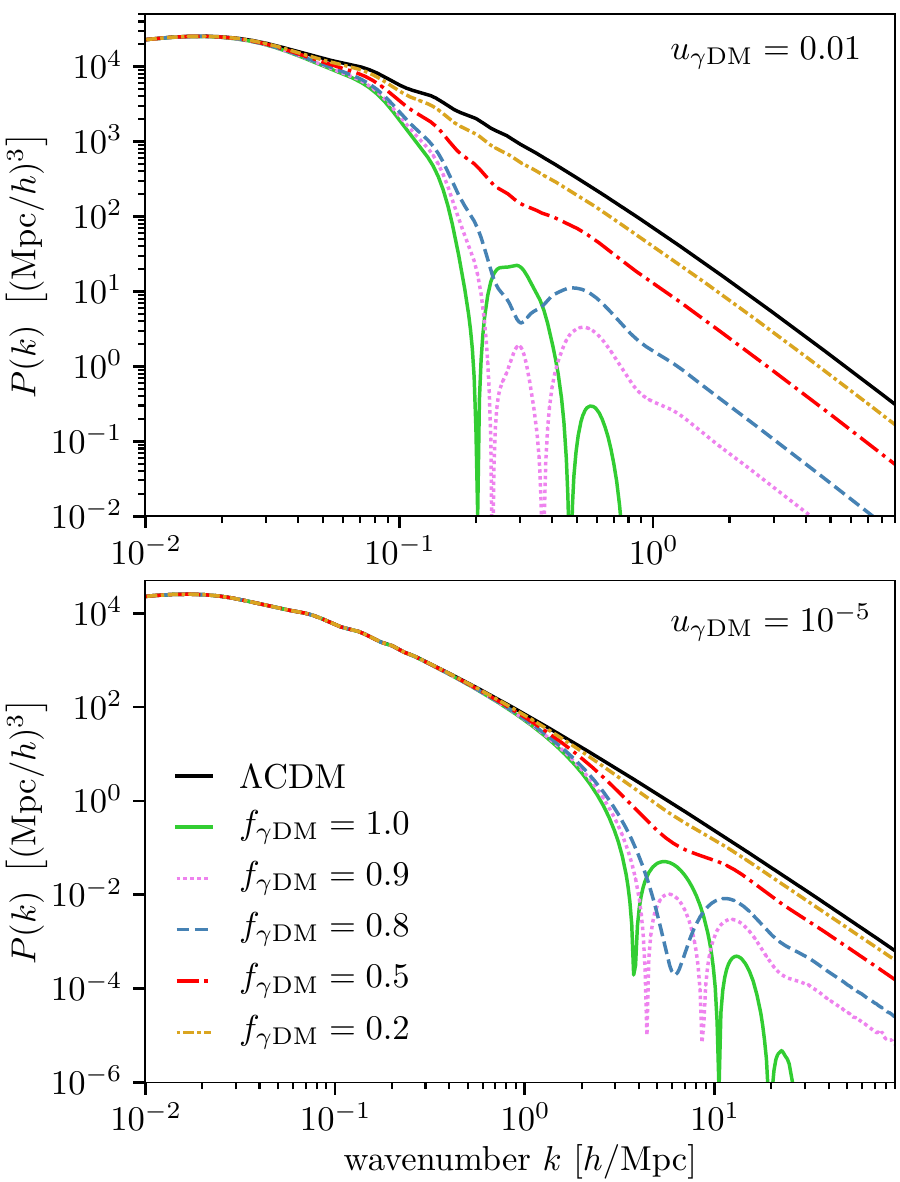}
\caption{The matter power spectrum for a cross section to mass ratio $\ugc=0.01$ (top) and $\ugc=10^{-5}$ (bottom) and different fractions of interacting DM.}
\label{fig: results-pk}
\end{figure}

The impact of mixed-DM on the linear matter power spectrum differs from that of pure-$\gdm$. As shown in Fig.~\ref{fig: results-pk}, the matter power spectrum of a pure-$\gdm$ scenario exhibits a series of damped (Bessel-like) oscillations at small-scales, while the linear matter power spectrum of mixed DM can be similar to the CDM spectrum in the presence of non-negligible neutrino masses. At large scales, however, there is no difference between  the \lcdm, the pure-$\gdm$, and the mixed-DM matter power spectrum. Figure~\ref{fig: results-pk} also shows that at intermediate scales, i.e. those corresponding to the exponential cut-off scale in the pure-$\gdm$ scenario, there is a suppression of power which can be more pronounced in the mixed-DM scenario than in the pure-$\gdm$ case. At small scales, the mixed-DM power spectrum evolves parallel to the \lcdm\  one but with a smaller amplitude, set by the fraction of interacting DM, $f_\gdm$. This classification holds true regardless of the precise value of $\ugc$: the cross section to mass ratio controls the scale at which the transition between the three regions occurs. We carefully explore now the three different regimes described above.

Firstly, the largest scales do not enter the Hubble radius until DM-photon interactions have kinetically decoupled, and therefore are not affected by the scattering processes. The scale factor of DM kinetic decoupling, $a_{\mathrm{DM},\mathrm{dec.}}$,  is determined by the condition
\begin{equation}
\ha \left(a_{\mathrm{DM},\mathrm{dec.}}\right) = 
\left.\frac{4\,\rho_\gamma}{3\,\rho_\gdm}\, a\, n_\gdm\,\sigma_\gdm
\right|_{a = a_{\mathrm{DM},\mathrm{dec.}}}
\,.
\end{equation}
Because the energy density of non-relativistic DM is proportional to its number density, the Hubble rate at decoupling only depends on the photon energy density. In the matter power spectrum, the scale at which the suppression due to collisional damping sets in is entirely governed by $u_\gdm$ but completely insensitive to $f_\gdm$.

\begin{figure}[t]
\includegraphics[]{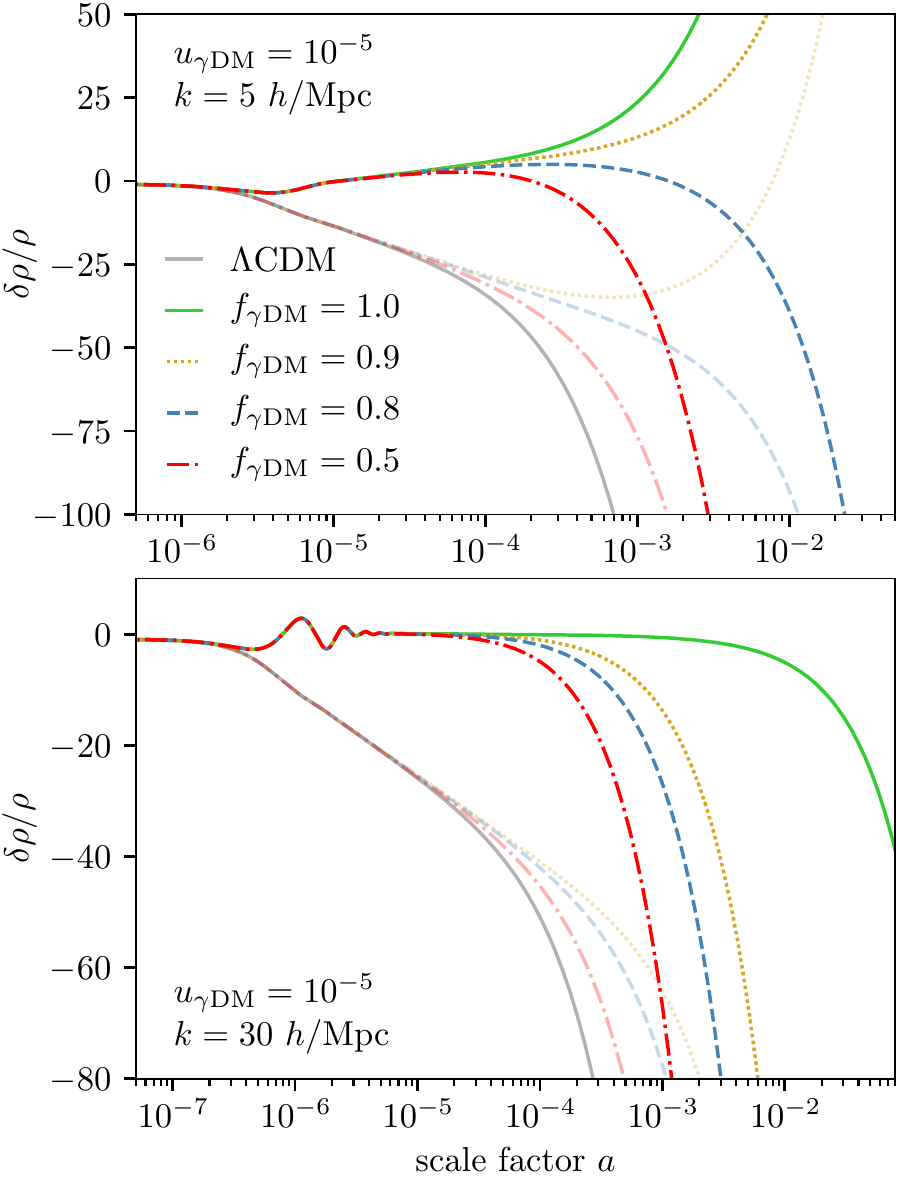}
\caption{Time evolution of two modes from the bottom panel of  Fig.~\ref{fig: results-pk}. The upper panel illustrates a mode with $k=5~h/\mathrm{Mpc}$, corresponding to the dip location in the mixed-DM matter power spectrum. The lower panel refers to a mode with $k=30~h/\mathrm{Mpc}$, corresponding to the regime  where the mixed DM matter power spectrum is parallel to that of \lcdm. The perturbations of the $\gdm$ (CDM) component are displayed in dark (light) colours.}
\label{fig: results-mode-evolution}
\end{figure}

Intermediate and small scales are more complicated to understand. Figure~\ref{fig: results-mode-evolution} shows the time evolution of two specific modes for $u_\gdm = 10^{-5}$ and varying fractions of interacting DM. The former ($k= 5 \, h/\rm{Mpc}$) lies at intermediate scales, precisely at the dip in the mixed DM matter power spectrum. The latter, larger mode ($k= 30 \, h/\rm{Mpc}$) is in the tail, where mixed DM and $\Lambda$CDM matter power spectra evolve parallel.

We first focus on intermediate scales, represented by the $k= 5 \, h/\rm{Mpc}$ mode, see the upper panel of Fig.~\ref{fig: results-mode-evolution}. Upon horizon entry, the density contrast $\delta\rho/\rho$ of the CDM component decreases with time, while the $\gdm$ component participates in the oscillations of the baryon-photon plasma for a short while, before the density contrast starts to grow.
Nevertheless, it is the absolute value of $\delta\rho/\rho$ what is important for the matter power spectrum.
For values of $u_\gdm \gtrsim  0.01$,  modes in the intermediate regime enter the Hubble radius before matter radiation equality ($a_\mathrm{eq} \simeq 3.0\times10^{-4}$). As the universe enters into the matter domination era, the growth of density perturbations increases, with an overall density contrast given by
\begin{equation}
\delta\rho \simeq \rho_\mathrm{DM} \left[f_\gdm\delta_\gdm + \left(1-f_\gdm\right)\delta_\mathrm{CDM}\right]  + \rho_\mathrm{b}\delta_\mathrm{b}\,,
\end{equation}
where $\rho_\mathrm{DM}$ is the combined $\gdm$ and CDM energy density and the subscript "b" refers to baryons. Two factors are decisive for the late time evolution of perturbations: the elapsed time between Hubble crossing and kinetic decoupling of the $\gdm$ perturbation, and the fraction of interacting DM. If the time that the $\gdm$ component spends in the coupled regime is short and simultaneously there is a significant fraction of interacting DM, the $\gdm$ component dominates the potentials and eventually determines the evolution of the DM perturbations. This is precisely what happens for the $f_\gdm = 0.9$ case illustrated in the top panel of Fig.~\ref{fig: results-mode-evolution}: after matter-radiation equality, the collisionless component turns around and follows the collisional one, i.e. eventually it grows in the positive direction as well. Because perturbations in the $\gdm$ component are damped initially and CDM perturbations start growing upon horizon entry,  the collisional component determines the evolution for a comparably large fraction of interacting DM. In these cases, where the metric evolution is dominated by CDM perturbations, the collisional component experiences the turn around after matter-radiation equality. The larger the fraction of collisionless DM, the earlier this turn around sets in. In any case, regardless of which DM species eventually dominates the evolution, the growth of perturbations is hampered while the collisional and the collisionless component compete, and this causes an additional power suppression in the mixed DM scenario on intermediate scales.

For an intuitive understanding how the competition between the two DM components hinders the growth of perturbations, it is useful to consider the configuration in position space. There, the sign difference between the CDM and the $\gdm$ perturbations corresponds to a configuration in which overdensities in the collisional component predominantly coincide with underdense regions in the collisionless component and vice versa. Hence, perturbations in the individual DM species partially chancel each other, and the potential wells are less deep than they would be for a single component DM of either kind. This slows down the growth of structures.

The cancellation between the two DM species is less severe not only if the fraction of interacting DM is small, but also if perturbations in the $\gdm$ component are sufficiently suppressed in comparison with the CDM component. The latter is the case on small scales, which cross the Hubble radius earlier, and where the collisional component is coupled to photons for a longer period. By the time the pressure from photon interactions ceases, perturbations in the CDM component are already well developed, and the $\gdm$ component falls in the potential wells created by the collisionless component. Some examples of this are shown in the bottom panel of Fig.~\ref{fig: results-mode-evolution}. Here, the $\gdm$ component follows the collisionless evolution regardless of the fraction of interacting DM. Still, because the potential wells are less deep during the initial phase of the growth of perturbations, there also is a suppression in the matter power spectrum on small scales. 

\subsection{Future constraints from LSS observations: forecasts for DESI}
\label{sec:desi}
\begin{figure}
\includegraphics[]{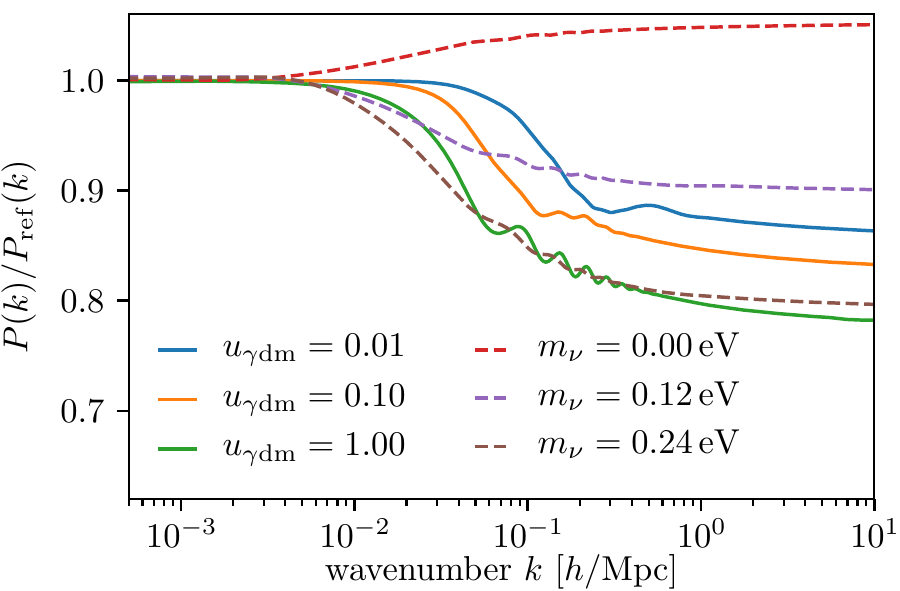}
\caption{The impact of heavy neutrinos and of mixed DM on the matter power spectrum whereby the reference scenario assumes $f_\gdm = 0$ and $m_\nu = 0.06\,\mathrm{eV}$ and all scenarios of mixed DM have $f_\gdm = 5\%$.}
\label{fig: pk-suppression}
\end{figure}
There is a striking similarity between the mixed DM scenario and a $\Lambda$CDM cosmology with massive neutrinos. In either case, some fraction of the late time DM energy density arises from a component which does not behave as a collisionless, cold fluid at early times. However, while the fraction of interacting DM, $f_\gdm$, can a priori take any value between zero and one, existing constraints on neutrino masses limit their fractional contribution to the matter density to lie within the $0.005 - 0.01$ range. In particular for a small interacting DM fraction, the effect on the matter power spectrum can be very similar to that of massive neutrinos, as Fig.~\ref{fig: pk-suppression} illustrates. Given their similarities, there is a risk of confusing mixed DM and massive neutrinos in the analysis of large scale structure data. In the following section we further illustrate this possibility with a Fisher forecast for the DESI survey \cite{Aghamousa:2016zmz}. DESI is expected to see first light in January 2020 \cite{Vargas-Magana:2019fvd} and can achieve an accuracy of $0.02\,\mathrm{eV}$ on the neutrino mass scale \cite{Font-Ribera:2013rwa} within a $\Lambda$CDM cosmology.

\begin{table}[!htbp]
\begin{center}
\begin{tabular}{c||c|c}
parameter & fiducial value & 1 $\sigma$ error\\
\hline
 & & \\[-8pt]
$\Omega_\mathrm{b}h^2$ & $0.022383$ & $1.5\times 10^{-4}$
\\
$\Omega_\mathrm{DM}h^2$ & $0.12011$ & $1.3\times 10^{-3}$
\\
$100\,\theta_\mathrm{s}$ & $1.040909$ & $3.2\times 10^{-4}$
\\
$n_\mathrm{s}$ & $0.96605$ & $4.3\times10^{-3}$
\\
$\ln\left(10^{10}\,A_\mathrm{s}\right)$ & $ 3.0488$ & $1.5\times 10^{-2}$ 
\end{tabular}
\end{center}
\caption{Five baseline cosmological parameters together with their fiducial values and Planck priors.}
\label{tab: basline-fiducial-parameters}
\end{table}

Our Fisher forecast proceeds in three steps. Firstly, we consider the DESI sensitivity to purely interacting DM, that is, we vary $u_\gdm$ but fix $f_\gdm =1$. Secondly, we investigate the mixed DM scenario in which both $u_\gdm$ and $f_\gdm$ are free parameters. In either of these cases, the neutrino sector consists of two massless and one massive neutrino species with mass $m_\nu = 0.06\,\mathrm{eV}$, and we have $N_\mathrm{eff} = 3.046$. Finally, we also allow $m_\nu$ to vary. In any of these three scenarios, there are five additional free baseline parameters, see Tab.~\ref{tab: basline-fiducial-parameters} for their fiducial values, for which we assume the Planck 2018 best-fit results~\cite{Aghanim:2018eyx}. The optical depth to reionization is kept fixed at $\tau_\mathrm{reio} = 0.0543$ and we consider modes up to $k_\mathrm{max} = 0.2\,h\,\mathrm{Mpc}^{-1}$.

DESI observes three different tracers of large scale structure, which are emission line galaxies (ELG), luminous red galaxies (LRG) and high-redshift quasars (QSO). We assume a linear bias model, that means the real-space linear matter power spectrum, $P_\mathrm{m}$, is related to the linear redshift-space galaxy power spectrum, $P_\mathrm{gg}$, by
\begin{equation}
P_\mathrm{gg}\left(k\right) = P_\mathrm{m}\left(k\right)
\times
\left(b + \beta\,\mu^2\right)^2\,,
\end{equation}
where $\mu$ is the angle between the mode $\mathbf{k}$ and the line of sight, $\beta$ is the growth rate
and $b$ the bias, relating the tracer's distribution to the DM distribution. Their individual Fisher matrices are combined using the multi-tracer approach of Refs.~\cite{Abramo:2013awa,Abramo:2011ph}. We use the values of the bias parameters from Ref.~\cite{Font-Ribera:2013rwa}, which are listed in Tab.~\ref{tab: linear-bias-parameters}. 
Here, $D(z)$ is the normalised growth factor. 
\begin{table}[!htbp]
\begin{center}
\begin{tabular}{c|c}
tracer & bias value \\
\hline \\[-.2cm]
emission line galaxies & $b_\mathrm{ELG}\left(z\right)\times D\left(z\right) = 0.84$
\\
luminous red galaxies & $b_\mathrm{LRG}\left(z\right)\times D\left(z\right) = 1.7$
\\
high redshift quasars & $b_\mathrm{QSO}\left(z\right)\times D\left(z\right) = 1.2$
\end{tabular}
\end{center}
\caption{Linear bias parameters for the tracers used here.}
\label{tab: linear-bias-parameters}
\end{table}

Although the matter power spectrum is sensitive to the effects of massive neutrinos and mixed DM, it can not constrain the six to eight free parameters of our scenarios by its own. We therefore add priors for the baseline parameters, based on the Planck results \cite{Aghanim:2018eyx}, in the form of a diagonal Fisher matrix with the squared Planck $1\sigma$ intervals (see Tab.~\ref{tab: basline-fiducial-parameters} for their values). This means we (conservatively) neglect any information that Planck data may provide on the neutrino masses or on the mixed DM scenario parameters, as well as possible cross-correlations. 

Before presenting the Fisher forecast results, a word of caution is needed here when interpreting our sensitivities. For non-Gaussian likelihoods the Cram{\'e}r-Rao inequality only provides a lower bound, while the error may be larger. Further, we estimate the Fisher matrix of each tracer accordingly to the well-known expression \cite{Tegmark:1997rp},
\begin{equation}
F_{ij} = \int_{-1}^{1} \int_{k_\mathrm{min}}^{k_\mathrm{max}} \frac{2\pi k^2\,dk\,  d\mu}{2\left(2\pi\right)^3}~
\frac{\partial \ln P_\mathrm{gg}}{\partial \theta_i}
\frac{\partial \ln P_\mathrm{gg}}{\partial \theta_j}~
V_\mathrm{eff}\left(k,\mu\right)\,,
\label{eq: fisher-pk}
\end{equation}
where
\begin{equation}
V_\mathrm{eff}\left(k,\mu\right) = \left(\frac{n\,P\left(k,\mu\right)}{1 + n\,P\left(k,\mu\right)}\right)\,V_\mathrm{survey}~,
\end{equation}
\noindent 
$n$ is the average number density of galaxies and $V_\mathrm{survey}$ the survey volume. Importantly, Eq.~(\ref{eq: fisher-pk}) was derived under the assumption of a Gaussian likelihood. Thus, the forecasted errors obtained for poorly constrained parameters, where the likelihood often will be non-Gaussian, should not be regarded as the realistic  but as the optimal ones. Such is the case, for instance, of the contours derived in the ($u_{\gdm}$, $f_{\gdm}$) parameter space. We nevertheless believe that the use of the Fisher approach is still valid, as it provides a straightforward method to qualitatively highlight the major difficulties and the strong parameter degeneracies in the analysis of models with mixed DM.

\subsection{Pure $\gamma$DM sensitivity}
\label{sec: gDM-sensitivity}
\begin{table}[!htbp]
\begin{center}
\begin{tabular}{c||c}
$u_{\gdm}$& $\delta\, u_{\gdm}$\\
\hline\\[-15pt] &   \\[-5pt]
$2.0\times 10^{-4}$ & $8.49\times 10^{-6}$ \\
$2.0\times 10^{-5}$ & $7.12\times 10^{-6}$
\end{tabular}
\end{center}
\caption{1-$\sigma$ marginalised error on the photon-DM interaction for different fiducial models from DESI plus Planck 2018 CMB priors.}
\label{tab: predictions-gdm}
\end{table}

In this first scenario, we fix the neutrino mass to $0.06$~eV and investigate the sensitivity of DESI to interacting DM, i.e. assuming that all the DM in the universe scatters elastically off photons with an interaction rate $u_{\gdm}$. The results for two possible fiducial values of $u_{\gdm}$ are summarised in Tab.~\ref{tab: predictions-gdm}.

Current limits on a pure interacting $\gamma$DM scenario from the analysis of Planck 2015 data establish that $u_{\gdm} \leq 2.3 \times 10^{-4}$ at 95\% CL for the most conservative TT+low TEB dataset, see Ref.~\cite{Stadler:2018jin}. Our results show that future matter power spectrum measurements from DESI, using different tracers, could improve this limit by more than one order of magnitude.

\subsection{Mixed $\gamma$DM and CDM sensitivity}
\label{sec: mixedDM-sensitivity}
\begin{table}
\begin{center}
\begin{tabular}{c|c||c|c}
$u_{\gdm}$ & 
$f_{\gdm}$ &
$\delta\, u_{\gdm}$ &  
$\delta\, f_{\gdm}$ \\
\hline\\[-15pt] & & &  \\[-5pt]
$1.0\times 10^{-3}$ & $0.1$ & $1.77\times 10^{-3}$ & $0.26$\\
$1.0\times 10^{-3}$ & $0.5$ & $3.72\times 10^{-4}$ & $0.27$\\
$1.0\times 10^{-2}$ & $0.05$ & $2.65\times 10^{-3}$ & $8.7\times 10^{-3}$
\end{tabular}
\end{center}
\caption{1-$\sigma$ marginalised error on the fraction of interacting DM and on its interaction rate with photons for different fiducial models from DESI plus Planck 2018 CMB priors.}
\label{tab: predictions-mdm}
\end{table}

In the mixed DM scenario, we have two extra parameters in addition to those listed in Tab.~\ref{tab: basline-fiducial-parameters}: the interaction strength $u_{\gdm}$ and the fraction of interacting DM $f_{\gdm}$. The results are shown in Tab.~\ref{tab: predictions-mdm} for different fiducial cosmologies satisfying the CMB limits derived in Sec.~\ref{sec:impact-cmb}. If a significant fraction of DM belongs to the interacting component whose cross section correspondingly is small, there is a strong degeneracy between $u_{\gdm}$ and $f_{\gdm}$ and the sensitivity of DESI to detect mixed DM reduces significantly in comparison to the purely interacting scenario. In contrast, for those regions of the parameter space where the CMB provides weak limits, i.e. for small values of $f_\gdm$ and comparatively large cross sections, the degeneracy between the mixed DM parameters is less severe. The pattern remains qualitatively similar in scenarios with varying neutrino masses and can be observed in the bottom left panels of Fig.~\ref{fig:mcmc}. In the latter case, departures from the $\Lambda$CDM power spectrum are less severe but arise on scales well above the non-linear cut-off we imposed for our analysis. Then, DESI is sensitive to the precise shape of the matter power spectrum. Importantly, large scale structure can probe the mixed DM scenario precisely in those regions of parameter space where CMB constraints are weakest.

\subsection{Combined sensitivity to mixed DM and neutrino masses}
\label{sec: mixedDM-mNU-sensitivity}
\begin{table*}[!htbp]
	\begin{center}
		\begin{tabular}{cc||ccc}
			$u_{\gdm}$ & 
			$f_{\gdm}$ &
			$\delta\, u_{\gdm}$ &  
			$\delta\, f_{\gdm}$ &
			$\delta\,m_\nu$\\
			\hline\\[-15pt] & & &  \\[-5pt]
			$0.1$ & $0.02$ & 
			$3.56\times 10^{-2}$ & $1.04\times 10^{-2}$ & $2.52\times 10^{-2}$
			\\
			$1.0\times 10^{-2}$ & $0.05$ &
			$2.95\times 10^{-3}$ & $8.70\times 10^{-3}$ & $2.08\times 10^{-2}$
			\\
			$1.0\times 10^{-3}$ & $0.5$ &
			$4.14\times 10^{-4}$ & $0.30$ & $2.08\times 10^{-2}$
			\\
			$1.0\times 10^{-3}$ & $0.3$ & 
			$6.79\times 10^{-4}$ & $0.30$ & $2.03\times10^{-2}$
			\\
			$4.0\times10^{-4}$ & $0.5$ & 
			$5.55\times 10^{-4}$ & $1.15$ & $2.05\times 10^{-2}$
		\end{tabular}
		\caption{1-$\sigma$ marginalised error on the fraction of interacting DM, its interaction rate with photons and the total neutrino mass for different fiducial models from DESI plus Planck 2018 CMB priors. The fiducial value for $m_\nu=0.06$~eV.}
		\label{tab: predictions-mdm-mnu}
	\end{center}
\end{table*}

\begin{figure*}
	\centering
	\begin{subfigure}[b]{0.5\textwidth}
		\centering
		\includegraphics[]{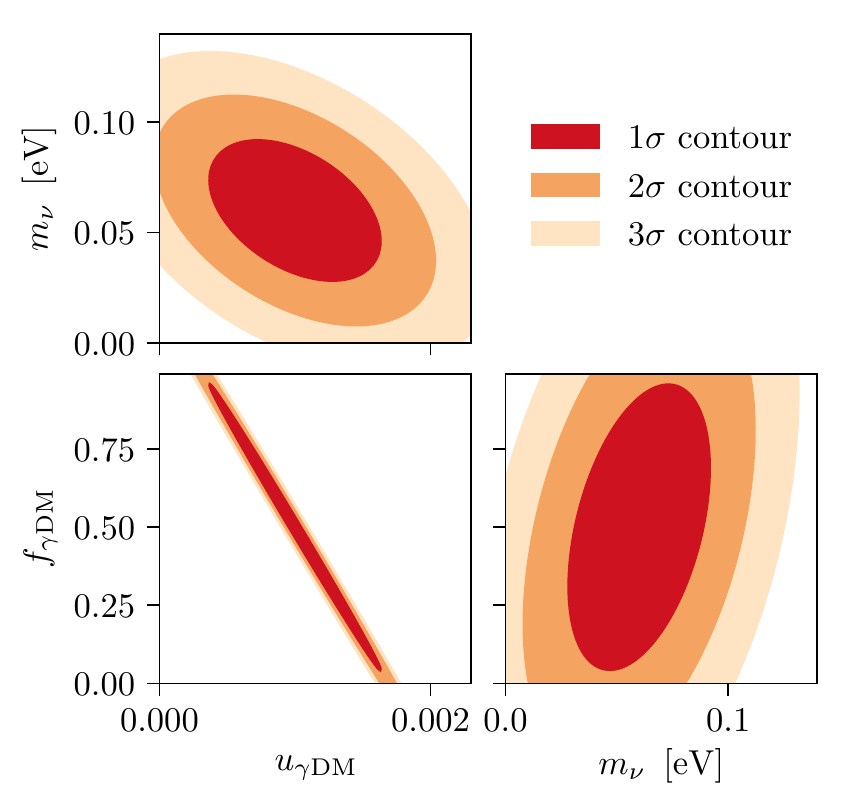}
		\caption{$u_{\gdm}=1.0\times 10^{-3}$, $f_{\gdm}=0.5$ and $m_\nu = 0.06\,\mathrm{eV}$.}
	\end{subfigure}%
	\begin{subfigure}[b]{0.5\textwidth}
		\centering
		\includegraphics[]{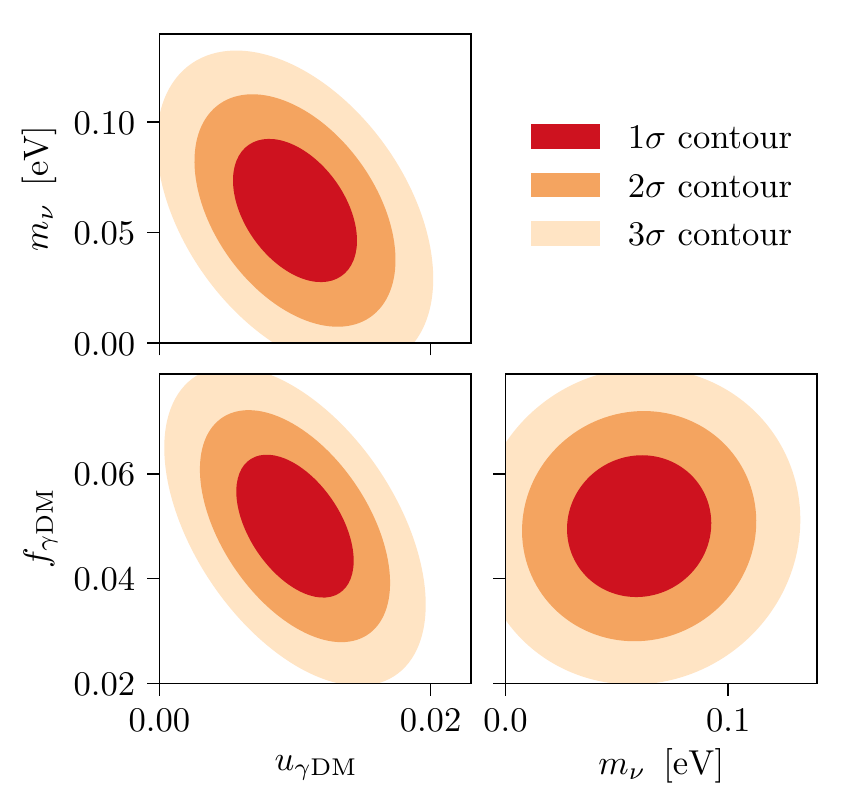}
		\caption{$u_{\gdm}=1.0\times 10^{-2}$, $f_{\gdm}=0.05$ and $m_\nu = 0.06\,\mathrm{eV}$.}
	\end{subfigure}
	\caption{Two-dimensional allowed contours for two fiducial scenarios.}
	\label{fig:deg}
\end{figure*}

In this last scenario, we investigate how well mixed DM and neutrino masses can be determined simultaneously from future DESI data. In the absence of any DM interactions, that is fixing $f_\gdm = u_\gdm=0$, we obtain a $1\sigma$ error on the neutrino mass of $\delta m_\nu \simeq 0.01\,\mathrm{eV}$. This value is smaller than the result quoted in Ref.~\cite{Font-Ribera:2013rwa}, most likely because we keep the optical depth to reionization fixed rather than treating it as a free parameter. However, it serves us as a baseline to which we can compare the expected limits on the neutrino mass in the presence of mixed DM. Table~\ref{tab: predictions-mdm-mnu} summarises our results and shows that the error on the neutrino mass at least doubles for all scenarios considered. The loss of sensitivity to the neutrino mass due to mixed DM thus is of similar order as that due to a modified dark energy sector \cite{Font-Ribera:2013rwa}. The two dimensional allowed contours in the ($u_{\gdm}$, $m_\nu$), ($u_{\gdm}$, $f_{\gdm}$) and ($m_\nu$, $f_\gdm$) planes, depicted in Fig.~\ref{fig:deg} for two benchmark cases, further illustrate the situation. For a large fraction of interacting DM and a relatively small interaction rate ($u_{\gdm}=1.0\times 10^{-3}$, $f_{\gdm}=0.5$) we observe a moderate degeneracy between $u_\gdm$ and $m_\nu$. A non-zero neutrino mass suppress the galaxy power spectrum at small scales as does an interacting DM component with non-negligible $u_{\gdm}$. Further, the aforementioned strong degeneracy between the two mixed $\gamma$DM parameters, $u_{\gdm}$ and $f_{\gdm}$, is inherited in the ($m_\nu$, $f_\gdm$) plane and the combination of both effects causes the larger neutrino mass error. As already explained in context of the mixed DM scenario (see Sec.~\ref{sec: mixedDM-sensitivity}), the degeneracy between $u_\gdm$ and $f_\gdm$ is less severe in the scenario with little interacting DM and a larger cross section ($u_\gdm=10^{-2}$, $f_\gdm=0.05$) on the right hand side of Fig.~\ref{fig:deg}. At the same time, the correlation between the neutrino mass and $u_\gdm$ further increases. As Fig.~\ref{fig: pk-suppression} reveals, it is precisely in this region of the parameter space where the similarity between the power spectra of the mixed DM and the massive neutrino scenarios are the most striking. Correspondingly, the most extreme cases listed in Tab.~\ref{tab: predictions-mdm-mnu} see a further increase of the neutrino mass error.
 
\section{Conclusions}
\label{sec: conclusions}

In this manuscript, we investigate a scenario with two dark matter (DM) components, in which one is interacting with photons and the other one behaves as a canonical Cold Dark Matter (CDM) fluid. The imprints on the Cosmic Microwave Background (CMB) and on the matter power spectrum $P(k)$ within such mixed DM scenarios are carefully analysed. 

Our CMB constraints, based on the Planck 2015 data, reveal a strong degeneracy between the interaction rate $u_\gdm$ and the interacting DM fraction $f_\gdm$. If the interaction rate is small enough, almost any fraction of interacting DM is permitted while, for small enough $f_\gdm$, the limits on $u_\gdm$ weaken considerably. Remarkably, considering the mixed DM scenario only and no uncertainties on the neutrino masses, large scale structure data can provide constraints which are highly complementary to those derived from the CMB. In particular it strengthens limits in the low-$f_\gdm$, high-$u_\gdm$ region of the parameter space.However, a further complication arises when we consider that the neutrino mass scale is not known precisely. The mixed DM  power spectrum can be very similar to the CDM one plus non-zero neutrino masses. In both scenarios, there is first a suppression of power followed by a similar evolution to CDM (which translates into a $P(k)$ parallel  to that of CDM but somewhat reduced in magnitude). The magnitude of the suppression is controlled by the fraction of interacting DM, while its onset depends on $u_\gdm$. We also observe an additional dip at the scale where the suppression sets in, which is caused by the partial cancellation of perturbations between the collisional and the collisionless DM components. 

Our forecast for the future Dark Energy Instrument (DESI) galaxy survey show that the neutrino mass bound weakens by at least a factor of two when the possibility of mixed DM is considered in the analysis. In particular in that region of the parameter space where a small abundance of interacting DM coincides with a comparably large interaction rate the neutrino mass scale is vulnerable to being overestimated due to the presence of DM. As Fig.~\ref{fig: pk-suppression} illustrates, this is precisely where similarities in the heavy neutrino and the mixed DM matter power spectra are the most striking. At this point we want to stress again the weakness of Fisher forecasts concerning one sided parameter constraints. Our analysis clearly demonstrates the degeneracy between neutrino masses and the mixed DM scenario and identifies the relevant region of the parameter space. To derive robust, quantitative bounds on the combined mixed DM and massive neutrino sensitivity we advocate for a MCMC analysis of the CMB and galaxy survey likelihood, which should focus on the $f_\gdm < 0.1$, $u_\gdm > 10^{-2}$ region.

CMB spectral distortions, finally, offer a complementary probe of DM interactions \cite{Ali-Haimoud:2015pwa,Diacoumis:2017hff} and can potentially lift the degeneracy between neutrino masses and the parameter of the mixed DM model. If interacting with photons, DM acts as a heat sink, and the spectral distortions generated are of the same order as ratio of number densities between DM and photons. Current bounds, assuming $f_\gdm=1$, imply $u_\gdm \lesssim 10^{-7}$ for DM lighter than $m_\gdm < 0.1\,\mathrm{MeV}$, but a future experiment like PIXIE has the potential to extend the limit to $\mathcal{O}\left(\mathrm{GeV}\right)$ masses \cite{Ali-Haimoud:2015pwa}.  Naively scaling the constraint with the reduced number density of interacting DM, implied by $f_\gdm < 1$, suggests that spectral distortions could be sensitive to $u_\gdm$ as small as $10^{-5}$ even for 1\% of light, interacting DM. Thus they can probe that region of parameter space where mixed DM affects neutrino mass measurements the most. We believe that this opportunity deserves further, detailed study.

\section*{Acknowledgements}
We would like to thank E. Villa for useful discussions. 
This project was funded by the Horizon 2020 research and innovation program under the Marie Sklodowska-Curie grant agreement No 674896. JS thanks the University of Sydney for hospitality, where part of this work was done. OM is supported by the Spanish grants FPA2017-85985-P and SEV-2014-0398 of the MINECO and the European Union's Horizon 2020 research and innovation program under the grant agreements No.\ 690575 and 674896.

\bibliography{gcdm-CMB-paper}
\end{document}